\documentclass[apj,letterpaper]{emulateapj}
\usepackage{apjfonts}

\newcommand{\arcm}{\hbox{$^\prime$}}
\newcommand{\etal}{{\rm et al.}\thinspace}
\newcommand{\eg}{{\it e.g.\ }}

\newcommand{\ie}{{\it i.e.\ }}

\newcommand{\degree}{\hbox{$^\circ$}}
\newcommand{\rosat}{\emph{ROSAT}}
\newcommand{\chandra}{\emph{Chandra}}
\newcommand{\xmm}{\emph{XMM-Newton}}
\newcommand{\xmms}{\emph{XMM}}
\newcommand{\asca}{\emph{ASCA}}
\newcommand{\einstein}{\emph{Einstein}}
\newcommand{\arcs}{\hbox{$^{\prime\prime}$}}
\newcommand{\Lx}{\ensuremath{L_{\mathrm{X}}}}

\newcommand{\Zsol}{\ensuremath{Z_{\odot}}}

\newcommand{\Msol}{\ensuremath{M_{\odot}}}
\newcommand{\LB}{\ensuremath{L_{\mathrm{B}}}}
\newcommand{\LBsol}{\ensuremath{L_{B\odot}}}
\newcommand{\LK}{\ensuremath{L_{\mathrm{K}}}}
\newcommand{\LKsol}{\ensuremath{L_{K\odot}}}

\newcommand{\s}{\ensuremath{\mbox{~s}}}
\newcommand{\ps}{\ensuremath{\s^{-1}}}
\newcommand{\cm}{\ensuremath{\mbox{~cm}}}
\newcommand{\pcmsq}{\ensuremath{\cm^{-2}}}
\newcommand{\pcmcu}{\ensuremath{\cm^{-3}}}
\newcommand{\km}{\ensuremath{\mbox{~km}}}
\newcommand{\Mpc}{\ensuremath{\mbox{~Mpc}}}
\newcommand{\pMpc}{\ensuremath{\Mpc^{-1}}}
\newcommand{\kmpspMpc}{\ensuremath{\km \ps \pMpc\,}}
\newcommand{\erg}{\ensuremath{\mbox{~erg}}}
\newcommand{\ergps}{\ensuremath{\erg \ps}}
\newcommand{\ergpspcmsq}{\ensuremath{\erg \ps \pcmsq}}
\newcommand{\kmps}{\ensuremath{\km \ps}}
\newcommand{\sas}{\textsc{sas}}

\newcommand{\Ho}{\ensuremath{H_\mathrm{0}}}

\newcommand{\Dtf}{\ensuremath{D_{\mathrm{25}}}}

\begin{document}

\title{On the Anomalous Temperature Distribution of the Intergalactic Medium
  in the NGC~3411 Group of Galaxies\footnotemark[*]} 

\author{E. O'Sullivan\altaffilmark{1}, J.~M. Vrtilek\altaffilmark{1},
  D.~E. Harris\altaffilmark{1} and T.~J. Ponman\altaffilmark{2}}
\altaffiltext{1}{Harvard-Smithsonian Center for Astrophysics, 60 Garden
  Street, Cambridge, MA 02138}
\altaffiltext{2}{School of Physics and Astronomy, The University of
  Birmingham, Birmingham, B15 2TT, UK}

\email{eosullivan@head.cfa.harvard.edu} 

\shorttitle{Anomalous Temperature Distribution in NGC~3411}
\shortauthors{O'Sullivan et al}

\footnotetext[*]{Based on observations obtained with XMM-Newton, an ESA science mission with instruments and contributions directly funded by ESA Member States and NASA}

\begin{abstract}
  We present \xmm, \chandra\ and VLA observations of the USGC~S152 group
  and its central elliptical NGC~3411. Imaging of the group X-ray halo
  suggests it is relaxed with little apparent structure. We investigate the
  temperature and metal abundance structure of the group halo, and find
  that while the abundance distribution is fairly typical, the temperature
  profile is highly unusual, showing a hot inner core surrounded by a cool
  shell of gas with a radius of $\sim$20-40~kpc, at the center of the
  larger group halo. Spectral mapping confirms an irregular ring of gas
  $\sim$0.15~keV cooler than its surroundings.  We estimate the total mass,
  entropy and cooling time profiles within $\sim$200~kpc, and find that the
  cool shell contains $\sim$9$\times10^9$\Msol of gas.  VLA observations at
  1.4, 5 and 8~GHz reveal a relatively weak nuclear radio source, with a
  core radio luminosity L$_{R}$=2.7$\times10^{38}$\ergps, and a diffuse
  component extended on scales of a few arcseconds (or more). A lack of
  evidence for activity at optical or X-ray wavelengths supports the
  conclusion that the central black hole is currently in a quiescent state.
  We discuss possible mechanisms for the formation of temperature features
  observed in the halo, including a previous period of AGN activity, and
  settling of material stripped from the halo of one of the other group
  member galaxies.
\end{abstract}


\section{Introduction}
\label{sec.intro}
The study of structure in the gaseous halos of galaxies, groups and
clusters has been greatly facilitated by the large collecting areas and
high spatial resolution of the \xmm\ and \chandra\ X-ray
observatories. Some useful comparisons of radio, optical and X-ray
structure were performed with earlier satellites
\citep[e.g.,][]{Sarazinetal95,McNamaraetal96}, but the relatively low
efficiency and/or broad point spread functions (PSFs) of the instruments on
\rosat, \asca\ and \einstein\ restricted this type of examination to the
brightest sources. \xmms\ and particularly \chandra\ have allowed a huge
expansion in the field, showing the range and variety of disturbed
structures, and demonstrating that they are relatively common. Perhaps most
notable are the detailed interactions between the radio jets of active
galaxies in the cores of groups and clusters and the surrounding
inter-galactic medium (IGM). Long \chandra\ observations of M87 in the
Virgo cluster \citep{Formanetal05,Churazovetal01} and NGC~1275 in Perseus
\citep{Fabianetal06,Sandersetal04,Fabianetal03b} have revealed complex features
indicative of shocks, bubbles of radio plasma, filaments of cooling gas and
sound wave propagation. Mergers and interactions between galaxies and
their environment also often produce visible structures in the IGM \citep[e.g.,][]{Kraftetal06,Mahdavietal05,Machaceketal05b}

In two prior papers \citep{OSullivanetal05,OSullivanetal03} we examined
\xmm\ observations of two poor, $\sim$2.5~keV clusters, AWM~4 and MKW~4.
Both were selected to be relaxed systems without significant structure, to
allow study of their radial mass and metal abundance profiles. The \xmms\ 
exposures revealed that while the two systems have regular surface
brightness distributions and similar total masses, they have distinctly
different temperature profiles. While MKW~4 shows a central temperature
decline (to $\sim$1.5~keV) similar to that seen in many groups and
clusters, AWM~4 is approximately isothermal despite a central cooling time
of only 5$\times10^8$ yr. This divergence appears to be related to the
active galactic nucleus (AGN) of NGC~6051, the dominant galaxy of AWM~4.
Spectral mapping shows regions of hot and high abundance gas in the inner
cluster halo, correlating with the AGN radio jets, suggesting that the halo
is being heated by the AGN activity. The differences between the two
clusters can therefore be attributed to their differing phase in the cycle
of cooling and AGN activity; If cooling is allowed to continue in MKW~4 it
is likely to fuel an AGN outburst in the central galaxy, while the current
activity in AWM~4 is likely to halt in the near future, as heating prevents
further gas cooling. The two clusters embody the extremes of the cycle,
undisturbed cooling and ongoing outburst. An interesting possibility
arising from this model is the observation of a system in the process of
changing state from cooling to heating.

NGC~3411 is the dominant, central elliptical of a galaxy group first
identified from the COSMOS catalogue of southern sky galaxies
\citep{Yentisetal92}. Later surveys have confirmed the existence of the
group, sometime referred to as SS2b~153 or USGC~S152, identifying three to
five bright galaxies as members
\citep{Ramellaetal02,Mahdavietal00,Giuricinetal00}. The velocity dispersion
of the group is estimated to be 211 \kmps, and the total mass of the system
$\sim$2.7$\times10^{13}$ \Msol. NGC~3411 is the only elliptical in the
group, and the other confirmed members are $\geq$1 mag fainter in the B
band. A short \rosat\ All-Sky Survey (RASS) observation showed the galaxy
to be at the center of an extended, regular X-ray halo \citep{Pierreetal94}
with a luminosity of $\sim6.5\times10^{43}$ \ergps\ \citep{Mahdavietal00}.
The galaxy was identified in the RASS Bright Source Catalogue
\citep{Vogesetal99}, and comparison with the NRAO\footnote{The National
  Radio Astronomy Observatory is a facility of the National Science
  Foundation operated under cooperative agreement by Associated
  Universities, Inc.} VLA Sky Survey \citep[NVSS,][]{Condonetal98} showed
that NGC~3411 also hosts a slightly extended radio source with a flux of
33~mJy at 1.4~GHz \citep{Baueretal00}. The galaxy has been observed as part
of the 6dF Galaxy Redshift Survey \citep{Jonesetal04}, and has an
absorption dominated optical spectrum, with no obvious signs of AGN
activity.

More recently, NGC~3411 has been observed by both \chandra\ and \xmm.
\citet{Vikhlininetal05} analysed the \chandra\ data as part of a sample of
13 nearby relaxed clusters, and briefly noted that it did not have a strong
central temperature decrement, perhaps indicating a recent AGN outburst or
stellar winds from the central elliptical. \citet{Mahdavietal05} included
the \xmms\ dataset as part of their study of eight RASSCALS groups and
found the group halo to be remarkably circular. They also noted that the
gas pressure within $\sim$10~kpc was a factor of 2 higher than expected
from self-similar scaling, possibly because of gas bound to the central
galaxy.

In this paper, we use \xmm, \chandra\ and Very Large Array (VLA)
observations to study the properties of the group and specifically to look
for signs of interaction between the central galaxy and the group X-ray
halo. Section~\ref{sec.Obs} describes the observations and their reduction
and section~\ref{sec.results} our analysis of the data and our main
results.  These are discussed in section~\ref{sec.discuss} and we summarise
our conclusions in section~\ref{sec.conc}. Throughout the paper we assume
\Ho=70 \kmpspMpc\ and normalize optical and near infra-red luminosities to
the absolute B and K band magnitudes of the sun, M$_{K\odot}$=3.37 and
M$_{B\odot}$=5.48 (equivalent to \LBsol=5.2$\times10^{32}$ \ergps).
Abundances are measured relative to the abundance ratios of
\citet{GrevesseSauval98}. Some details of the location and scale of
NGC~3411 are given in Table~\ref{tab.basics}.

\begin{deluxetable}{lr}
\tablewidth{0pt} 
\tablecaption{\label{tab.basics} Location and scale of NGC~3411.}  
\tablehead{} 
\startdata
R.A. (J2000) & 10$^h$50$^m$26.1$^s$ \\
Dec. (J2000) & -12$^d$50$^m$42.3$^s$ \\
Redshift & 0.01527 \\
Distance (\Ho=70) & 65.4 \\
1 arcmin = & 19~kpc \\
\Dtf\ radius = & 19.4~kpc\\
log \LB\ (\LBsol) & 10.66 \\
log \LK\ (\LKsol) & 11.41 \\
log \Lx\ (\ergps) & 42.51$^{+0.03}_{-0.02}$ \\[-2mm]
\enddata 
\tablecomments{\Lx\ is the unabsorbed 0.4-7.0~keV X-ray luminosity within
  662.5\arcs\ (210~kpc), taken from our own spectral fits}
\end{deluxetable}

\section{Observations and Data Analysis}
\label{sec.Obs}

\subsection{\xmm\ data}
The NGC~3411 group was observed with \xmm\ during orbit 555 (2002 December
20-21) for just under 36,000 seconds (ObsID 0146510301). The EPIC
instruments were operated in full frame mode, with the thin optical
blocking filter. A detailed summary of the \xmm\ mission and
instrumentation can be found in \citet[and references
therein]{Jansenetal01}. Reduction and analysis were performed using
techniques similar to those described in \citet{OSullivanetal05}.  The raw
data from the EPIC instruments were processed with the most recent version
of the \xmm\ Science Analysis System (\textsc{sas v.6.5.0}), using the
\textsc{epchain} and \textsc{emchain} tasks. Bad pixels and columns were
identified and removed, and the events lists filtered to include only those
events with FLAG = 0 and patterns 0-12 (for the MOS cameras) or 0-4 (for
the PN). The total count rate for the field revealed a short background
flare in the second half of the observation. Times when the total count
rate deviated from the mean by more than 3$\sigma$ were therefore excluded.
The effective exposure times for the MOS and PN cameras were 22.2 and 19.2
ksec respectively.

Images and spectra were extracted from the cleaned events lists with the
\sas\ task \textsc{evselect}. Response files were generated using the \sas\ 
tasks \textsc{rmfgen} and \textsc{arfgen}. Because the group is relatively
X-ray bright, and the PN detector mode uses a short frame integration time
(73 ms), there are a significant number of out-of-time (OOT) events,
visible in the PN data as a trail extending from the center of the source
toward the CCD readout. An OOT events list was created using
\textsc{epchain}, and scaled by 0.063 to allow statistical subtraction of
the OOT events from spectra and images.  Point sources were identified
using a sliding cell algorithm, and all data within 17\arcs\ of point
sources were removed.

Creation of background images and spectra for the system was hampered by
the fact that the group X-ray halo extends beyond the field of view. Use of
the ``double-subtraction'' technique \citep{Arnaudetal02,Prattetal01}
involves correcting blank-sky exposures to match a source-free area of the
observation; for NGC~3411 this is not feasible. We therefore decided to
consider the background as two separate components. The background
contribution from high-energy particles can be estimated based on events
which are detected outside the field of view. Particles-only datasets, made
up of exposures taken with the telescope shutters closed
\citep{Martyetal03}, were scaled to match these events and used as basis of
the background. The particle component dominates the continuum at high
energies and produces a number of strong spectral lines, which vary across
the field of view. The particles-only datasets contain these features, and
subtraction of the scaled data from regions matching those used in the
source data removes them satisfactorily.

The low energy background arises primarily from soft X-rays emitted by hot
gas in the galaxy, and from coronal emission associated with solar wind
interactions. This component can be modelled as a cool plasma extending
uniformly over the field of view. Trial fitting of spectra extracted in the
outer part of the field of view, and of a \rosat\ All-Sky Survey spectrum
extracted from an annulus at 1.5-2.0\degree\ from the group center showed
that a 0.2 keV, solar abundance plasma model provided an acceptable fit.
The variation of the soft contribution across the field of view can then be
calculated based on a soft energy exposure map, or via the spectral
response matrices. We add this plasma model as an additional component in
our spectral fits, with all components frozen. For surface brightness
fitting, an exposure map image scaled to match the plasma model
contribution is added to the particles-only background image. We note that
this may be a rather simple model for the fairly complex soft background,
and ignores issues of small scale variation in the background. However, the
group halo is relatively bright and contributes the majority of soft counts
at all radii, so variations in the soft background have a relatively
minimal impact.

\subsection{\chandra\ data}
NGC~3411 was observed with the ACIS instrument during \chandra\ Cycle 3, on
2002 November 5-6 (Obs ID 3243). A summary of the \chandra\ mission and
instrumentation can be found in \citet{Weisskopfetal02}. The S3 chip was
placed at the focus of the telescope in order to take advantage of the
enhanced sensitivity of the back illuminated CCDs at low energies. The
instrument operated in very faint mode, and observed the target for just
under 30 ksec. Reduction and analysis was performed using methods similar
to those described in \citet{OSullivanPonman04b}.  The level 1 events
file was reprocessed using \textsc{ciao} v3.2 and \textsc{caldb} v3.2.0.
Bad pixels and events with \asca\ grades 1, 5 and 7 were removed. The data
were corrected to the appropriate gain map, the standard time-dependent
gain correction was made, and a background light curve was produced.
Background flaring occurred through the first early part of the
observation, and all periods where the count rate deviated from the mean by
more than 3$\sigma$ were excluded, so as to prevent contamination. The
exposure time of the observation after cleaning was 18.3 ksec. We chose to
use only data from the S3 chip, as the decline in effective area and
broadening Point spread Function (PSF) on the more distant chips made data
extracted from them less useful than those available from \xmm.

Point sources were identified using the \textsc{ciao} task
\textsc{wavdetect}, with a detection threshold of 10$^{-6}$, chosen to
ensure that the task detects $\leq$1 false source in the field. Source
ellipses were generated with axes of length 4 times the standard deviation
of each source distribution, and regions of twice this size were
removed from the data before further analysis. A source was detected
coincident with the peak of the diffuse X-ray emission; this was considered
a potentially false detection and ignored, though we did later test for the
presence of a central X-ray point source. Images and spectra were extracted
from the resulting ``clean'' events list and appropriate weighted spectral
response matrices were generated using the \textsc{mkwarf} and \textsc{mkwrmf}.

Background estimation for spectral fitting was performed in an analogous
manner to that described above for the \xmm\ data. The blank-sky datasets
available as part of the \chandra\ \textsc{caldb} were used to subtract the
high energy background. Spectra extracted from the blank-sky events list,
scaled to match the numbers of counts in PHA channels 2500-3000 in the
source data, were used as the basis of the background. Examination of
spectra from various radii shows that the blank-sky data matches the high
energy continuum and instrumental line features in the source data well.  A
0.2 keV plasma model was included in the spectral fits to account for any
galactic emission not found in the blank-sky data, but testing showed that
this had little impact on the fits. The halo of NGC 3411 completely
dominates the soft emission within the S3 field of view.

\subsection{VLA data}
As mentioned in Section~\ref{sec.intro}, archival NVSS data shows a
slightly extended radio source coincident with the centre of NGC~3411. In
order to improve on the spatial resolution of these data, NGC~3411 was
further observed twice with the VLA. Details of the observations are given
in Table~\ref{tab.vlabasics}. The data were reduced in a standard fashion
using AIPS.  Because of the low intensity of the sources, no self-cal was
performed and no polarization maps were made.  The flux calibrator was
3C286 (observed twice during each run) and the phase calibrator was
J1039-156.  The final maps were made with IMAGR.  For the 1.4 GHz data, it
was necessary to make a tapered map of a large area to locate sources far
from the field center in order to include them in the final cleaning
process.  Flux densities were measured using IMEAN and IMFIT, the latter
task also providing deconvolved source sizes for two dimensional Gaussian
fits. The results of our analysis of the VLA data are discussed in Section~\ref{sec.radio}.

\setlength{\tabcolsep}{1mm}
\begin{deluxetable}{lccccccc}
\tablecaption{\label{tab.vlabasics} Summary of VLA observations.}  
\tablehead{\colhead{Freq.} & \colhead{Date} & \colhead{Array} &
  \colhead{t$_{exp}$} & \colhead{r.m.s.} & \colhead{Beam Size} &
  \colhead{P.A.} & \colhead{Program} \\ \colhead{(GHz)} & & & &
  \colhead{(mJy)}& \colhead{(\arcs)} & \colhead{(\degree)} & } 
\startdata
8.46 & 2005Sep04 & C & 1h6m & 0.029 & 4.1$\times$2.8 & 8.6 & AO196 \\
4.86 & 2006Feb15 & A & 3h1m & 0.016 & 0.66$\times$0.43 & 3.0 & AO202 \\
1.425 & 2006Feb15 & A & 2h24m & 0.044 & 2.3$\times$1.6 &-2.4 & AO202 
\enddata 
\end{deluxetable}

\section{Results}
\label{sec.results}
\begin{figure*}
\centerline{\includegraphics[angle=-90,width=\textwidth]{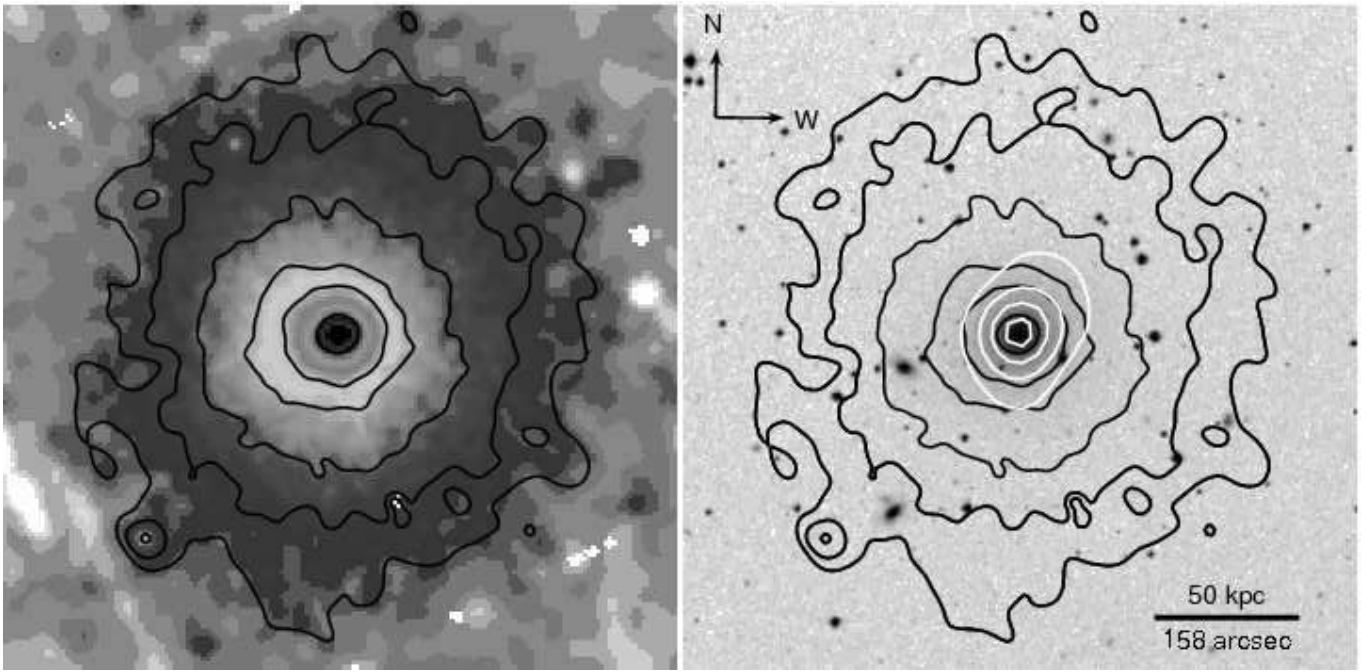}}
\caption{\label{fig.overlay}\textit{Left}: Adaptively smoothed mosaiced
  0.3-7~keV \xmm\ EPIC MOS+PN X-ray image of NGC~3411, with contours
  overlaid for clarity. Smoothing was performed with the \textsc{sas
    asmooth} task, requiring a signal-to-noise ratio of 10 in the smoothed
  image. \textit{Right}: DSS optical image of NGC~3411 with \xmms\ X-ray
  contours overlaid in black and NVSS radio contours in white. Both panels
  have the same scale and alignment. The NVSS contours begin at 5$\sigma$
  significance (1.5 mJy/beam) and increase in factors of 2. The X-ray
  contours are set at 0.65, 2, 6, 24, 48, 196 counts/pixel, summed over all
  three cameras.}
\end{figure*}

Figure~\ref{fig.overlay} shows a Digitized Sky Survey (DSS) optical image
of NGC~3411 with smoothed \xmm\ X-ray contours overlaid. The contours were
taken from an adaptively smoothed mosaiced image combining data from all
three \xmms\ EPIC cameras. Smoothing was performed with the \textsc{sas
  asmooth} task, with smoothing scales chosen to achieve a signal-to-noise
ratio of 10. The X-ray halo extends well beyond the field of this image,
and appears to be roughly circular at all radii. Examination of smoothed
and unsmoothed \xmms\ and \chandra\ images suggests that in general the
group halo relatively relaxed, with no major substructure. We see no areas
of enhanced or decreased surface brightness, or steep surface brightness
gradients, such as those associated with shocks, cavities or cold fronts.

\subsection{Surface brightness modeling}
\label{sec.sb}

\begin{figure*}
\centerline{\includegraphics[width=\textwidth,bbllx=10,bblly=210,bburx=562,bbury=730,clip=]{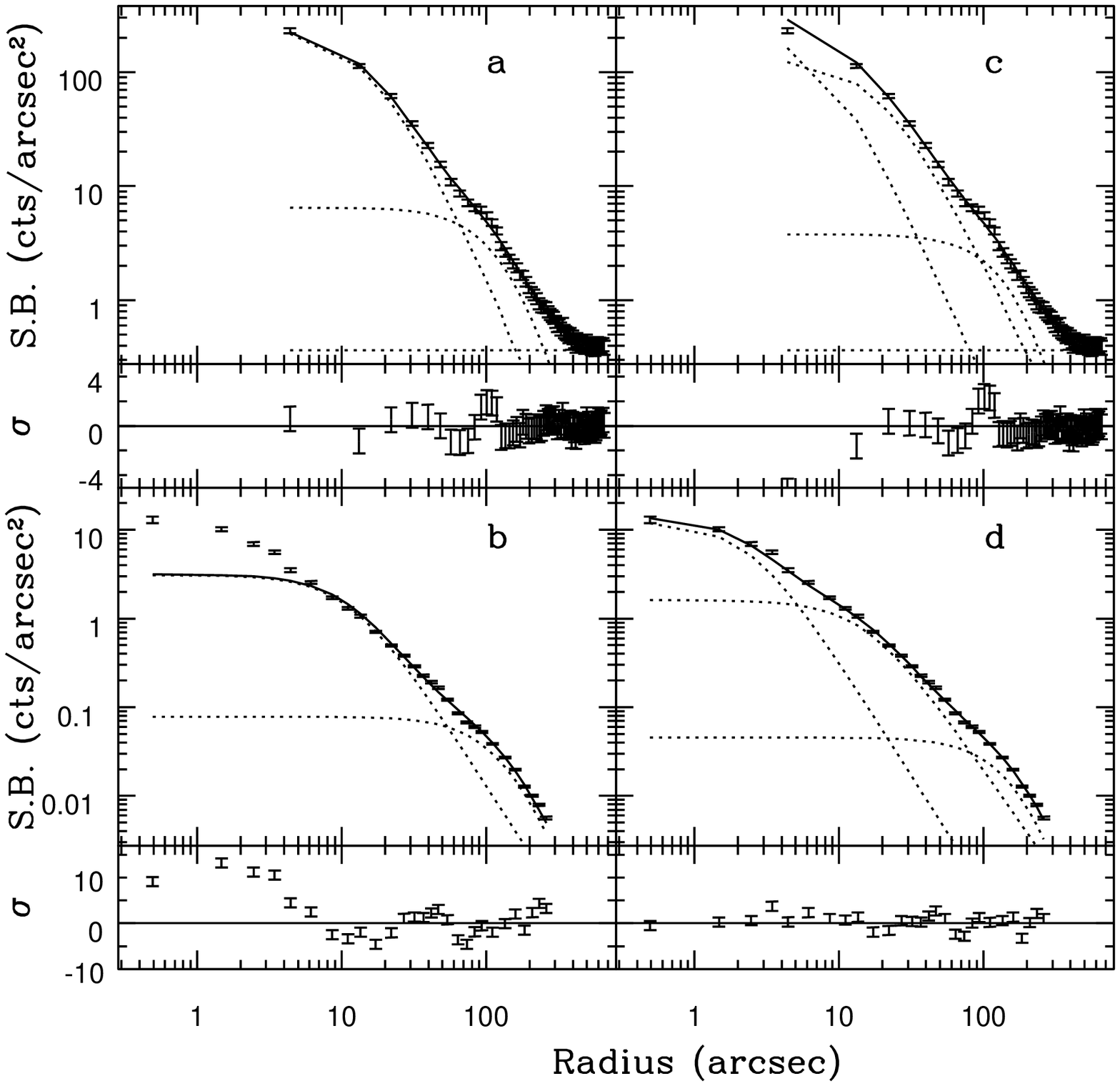}}
\caption{\label{fig.sb}
  Azimuthally averaged radial surface brightness profiles based on \xmm\ 
  (plots a and c) and \chandra\ (plots b and d) data, with residuals
  shown in terms of significance. Solid lines indicate
  surface brightness models consisting of two (plots a and b) or three
  (plots c and d) beta models. Dotted lines show individual model
  components. See Section~\ref{sec.sb} for further details.}
\end{figure*}

We initially fitted 1-dimensional azimuthally averaged radial surface
brightness profiles.  0.3-3.0 keV images were extracted for the EPIC
instruments and the ACIS-S3 chip, and appropriate exposure maps generated.
The \xmms\ images ere binned to the physical pixel scale of the PN camera
(4.4\arcs), and the \chandra\ image to the pixel size of the ACIS CCDs
(0.492\arcs). The PN image was corrected for OOT events. Radial profiles
were then extracted, using annuli of width 2 pixels (\ie, 8.8\arcs) for
\xmms\ and 2-60 pixels ($\sim$1-27\arcs), increasing with radius, for
\chandra.  The \xmms\ profiles extend to a radius of 11\arcm, the \chandra\
profile to $\sim$4.3\arcm. PSF images for PN and MOS cameras were extracted
from the calibration archive.  Prior experience has shown that PSF
convolution is necessary for accurate fitting of \xmm\ data, but has little
effect on \chandra\ surface brightness fits, owing to the very narrow
\chandra\ PSF. We therefore chose to include PSF convolution in the \xmms\
fitting, but excluded it from the \chandra\ fits.  The \xmm\ images were
summed to create a single radial profile as were profiles taken from the
PSF images. Using these data, we found that our background models and
exposure correction provided a good match to the data, to the limits of the
profiles. Figure~\ref{fig.sb} shows the resulting surface brightness
profiles.

Fitting was carried out in \textsc{ciao sherpa}, and tests were made
fitting the two profiles individually.  The \xmms\ profile clearly extends
far enough to reach the background level, which is modelled by a constant
component which is unnecessary in the \chandra\ fits. A single beta model
proved a poor fit to both profiles, with clear deviations from the data at
large and small radii, particularly in the \xmms\ data. Many groups have
surface brightness profiles which require more than one beta model to
produce a good fit \citep[e.g.][]{Helsdonponman00}. This usually indicates
an extra component of emission, for example a galaxy halo or cooling flow
embedded within a larger group halo.  We therefore fitted a second beta
model, which provided a visibly improved fit to the \xmms\ profile. The
second component provides a slight increase in surface brightness to match
a ``bulge'' in the profile at $\sim$100\arcs, and a somewhat steeper slope
beyond 300\arcs.  The model was a good fit to the \chandra\ profile outside
$\sim$10\arcs, but underestimated the central surface brightness
significantly (see panels a and c of Fig.~\ref{fig.sb}).
Table~\ref{tab.sb} gives details of this model, which was fitted to the
\xmms\ profile only (a goodness of fit for the \chandra\ data is given for
comparison, but the model was not fitted to the \chandra\ data).  The
residual emission is clearly more extended than would be expected for a
point source, and a model including a central point source did not provide
a good fit to the central region.

Addition of a third beta model did provide a good fit to the central region
of the \chandra\ profile, but simultaneous fitting of both datasets showed
that this model overestimated the central surface brightness of the \xmms\ 
profile by $\sim$25 per cent.  This disagreement seems most likely to be
related to the broad \xmms\ PSF, which will act to reduce the peak surface
brightness of the EPIC images.  However, the inclusion of PSF convolution
in the fit might be expected to correct this issue. We use monoenergetic (1
keV) PSF images to create our PSF model, and it is possible that this
introduces some error in the convolution, though we would not expect it to
be noticeable. It is also possible that the \textsc{sherpa} convolution
algorithm is imperfect, though testing shows that convolution of a delta
function with the PSF model reproduces the PSF to reasonable accuracy. In
order to test whether the 2-D PSF convolution is more accurate that 1-D, we
performed a 2-D fit to the \xmms\ data using a circular model based on the
\chandra\ 1-D fit with normalization free to vary. We again find that the
model overestimates the data, though in this case there is a possibility
that the slight ellipticity of the halo produces a fit which is a close
match to the data at moderate radii but poorer in the core. However, it
does not appear that fitting in 2-D resolves the problem, suggesting that
the 1-D PSF convolution is not causing the difference. The differing
energy responses of the telescopes and detectors should be accounted for by
the exposure maps. In any case, the difference is relatively minor,
affecting only the central \xmms\ bin.

\begin{deluxetable}{lcc}
\tablewidth{0pt}
\tablecaption{\label{tab.sb} Parameters of 1-D surface brightness models.}
\tablehead{\colhead{} & \colhead{2-$\beta$} & \colhead{3-$\beta$} }
\startdata
$R_{core,1}$ (\arcs) & 11.82$^{+2.57}_{-2.04}$ & 15.44$^{+4.01}_{-2.31}$ \\[1mm]
$R_{core,2}$ (\arcs) & 147.37$^{+66.56}_{-32.18}$ & 206.04$^{+47.92}_{-39.17}$ \\[1mm]
$R_{core,3}$ (\arcs) & - & 2.46$^{+0.80}_{-0.49}$ \\[1mm]
$\beta_{1}$  & 0.59$^{+0.09}_{-0.07}$ & 0.56$^{+0.07}_{-0.02}$ \\[1mm]
$\beta_{2}$  & 0.88$^{+0.41}_{-0.14}$ & 1.09$^{+0.31}_{-0.21}$ \\[1mm]
$\beta_{3}$  & - & 0.60$^{+0.23}_{-0.10}$ \\[1mm]
red. $\chi^2$ & (0.429/30.561) 7.307 & (0.983/3.934) 1.486 \\[1mm]
d.o.f. & (68/22) 96 & (66/19) 94 \\
\enddata
\tablecomments{Values for reduced $\chi^2$ and degrees of freedom of the
  fits are given in the format ``(\xmms/\chandra) total'', so as to allow
  comparison of the quality of fit between instruments. Note that the
  2-$\beta$ model was fitted only to the \xmms\ data, while the 3-$\beta$
  model was fitted to both datasets simultaneously.}
\end{deluxetable}

We also perform 2-dimensional image fitting to the two datasets, in order
to test for the presence and effects of elliptical components. Simultaneous
fits were performed using the three \xmms\ images described above, and
background images constructed using the method discussed in
Section~\ref{sec.Obs}. Exposure maps were used to model the instrumental
response, and models were convolved with the appropriate PSF images. As
numerous image pixels contained few or zero counts, we used the
\textsc{xspec} implementation of the Cash maximum likelihood statistic
\citep{Cash79} in \textsc{sherpa}. We note that this statistic does not
provide a true measure of the absolute goodness of fits, and we therefore
judge the quality of fits by inspection of radial profiles and residual
images, as well as the fit statistic.

Fitting to the \xmms\ images, we again find that a single beta model is an
inadequate description of the data, overestimating the central surface
brightness and underestimating it between at 80-150\arcs.  A two-beta model
provides a reasonable fit to the data. The centrally dominant component has
a slightly smaller core radius than that found in the 1-D fits, but a very
similar slope (R$_{core}$=6.91$^{+0.14}_{-1.755}$\arcs,
$\beta$=0.54$^{+0.01}_{-0.04}$, 1$\sigma$ uncertainties).  The outer
component is more divergent, with a smaller core radius
(R$_{core}$=85.59$^{+42.77}_{-4.57}$\arcs) and corresponding flatter slope
($\beta$=0.64$^{+0.17}_{-0.01}$). Both components are mildly elliptical
($\epsilon_{inner}$=0.19$^{+0.04}_{-0.08}$,
$\epsilon_{outer}$=0.15$^{+0.07}_{-0.15}$), though the outer component is
poorly constrained, and consistent with zero ellipticity. The two
components have slightly offset centers, but only by $\sim$2 image pixels,
well within the size of the PSF. In general the model appears to be a good
fit to the images, with the mild ellipticity trading off with the slope of
the outer component.

\subsection{Spectral fits}
\label{sec.spectra}
Most relaxed gaseous halos have temperature and abundance structures which
vary smoothly with radius. As NGC~3411 is roughly circular in projection,
circular annuli were used to extract spectra from the \xmms\ EPIC and
\chandra\ datasets. Appropriate response files were created, as well as
background spectra as described in Section~\ref{sec.Obs}. Spectral regions
were chosen to ensure at least 8000 source counts in each bin. As expected
given the differences in collecting area, this produced rather narrower
bins in the \xmm\ data than in \chandra. Data from the two satellites were
fitted independently, for this reason and to allow the two to act as a
check upon each other. Counts at energies lower than 0.4 keV or higher than
7.0 keV were ignored, as the low energy response of some of the instruments
is questionable, and energies higher than 7.0 keV tend to be dominated by
cosmic and particle background emission, and fluorescence lines within the
detectors. The hydrogen column of our models was held at the galactic value
(4.43$\times10^{20}$ cm$^{-2}$) in all fits. Spectra were grouped to 20
counts per bin. The spectra were initially examined and fit by hand using
\textsc{xspec} v11.3.2p and were found to be acceptably described by single
temperature APEC plasma models \citep{Smithetal01}.

\begin{figure}
\centerline{\includegraphics[width=\columnwidth,bbllx=40,bblly=210,bburx=570,bbury=750,clip=]{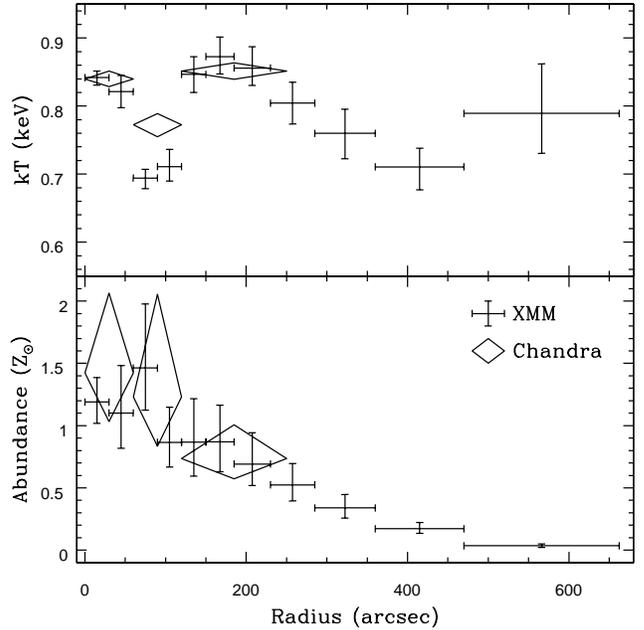}}
\caption{\label{fig.deproj}
Deprojected temperature and abundance of the group halo, from independent
\chandra\ and \xmms\ fits. 1$\sigma$ uncertainties are indicated.}
\end{figure}

As group halos are extended along the line of sight as well as in the plane
of the sky, and spectral properties are expected to vary with true rather
than projected radius from the group center, we perform a deprojection
analysis. Using two parallel scripts, one based on the \textsc{xspec
  projct} model and another written using \textsc{sherpa}, we fit the
\chandra\ and \xmm\ spectra and find the temperature and abundance profiles
shown in Figure~\ref{fig.deproj}.  The two scripts produced almost
identical results, suggesting that the fits are robust.  The abundance
profile is simple, declining from a central peak to low values at large
radii, as expected. The temperature profile is more complex.  Typically,
relaxed groups have temperature profiles with a central cool region, a peak
at moderate radii, and a decline to lower temperatures in the outer halo
\citep[see e.g., MKW~4,][]{OSullivanetal03}.  The NGC~3411 profile follows
this pattern at radii $>$60\arcs, but inside this radius the temperature is
higher than expected. The profile suggests that the group has a hot core,
surrounded by a shell of cool gas, in turn surrounded by hotter gas whose
temperature slowly declines with radius.

The radius of the cool shell, 60-120\arcs, corresponds well with the
``bulge'' seen in the surface brightness profiles in Section~\ref{sec.sb}.
We would expect the cooler gas of the shell to be slightly more luminous
than the warmer gas surrounding it if it is in pressure equilibrium, so
this is the likely cause of the excess surface brightness. While the
difference between minimum and maximum temperature is small, $\sim$0.15 keV
for the \xmm\ data, it is statistically significant; for the \chandra\ 
profile, the temperature difference between bins 1 and 2 is 3.4$\sigma$
significant, and between 2 and 3, 3.9$\sigma$ significant; for the \xmms\ 
profile, bins 2 and 3 are different at the 4.75$\sigma$ level and bins 4
and 5 are different at the 3.7$\sigma$ level.  It is notable that there is
a disagreement between the \chandra\ and \xmms\ profiles as to the
temperature of the cool shell. It seems likely that this arises due to the
rather broad bins of the \chandra\ profile and the lack of data points at
radii beyond 230\arcs. Deprojection uses the outer bins to estimate the
contribution of emission at large radii, in the line of sight, to spectra
at smaller projected radii, and then correct for it.  Inaccuracies in
fitting at large radii are therefore propagated inwards; if the fitted
temperature in a bin is too high, it is likely to lower the temperatures of
the bins at smaller radii. In this case, the outermost \chandra\ bin is
broad and probably contains some variation in temperature and abundance
which is then approximated, but not completely described, by the fitted
model. The bin will also contain some cool emission from outer shells which
may affect the fit, lowering the temperature. This could potentially lead
to an underestimate of the temperature in bin 3, an undersubtraction of
hard emission and therefore a higher temperature in bin 2. In any case, the
fact that bins 3 and 4 of the \xmms\ profile agree with one another
suggests that the disagreement is likely to be an issue with the \chandra\ 
analysis, and that the \xmms\ measurements are more likely to be correct.

To test the robustness of the temperature measurements in the inner region
we froze the contributions from the outer bins and fitted a range of models
instead of the simple single temperature plasma used in the deprojection.
For the two central \xmms\ bins, we tested two temperature plasma models,
plasma models with elemental abundances free to vary, and combinations of
plasma and powerlaw models. In both bins, two temperature plasma models
improve the fits, but do not lower the temperature of the main spectral
component. The secondary components have temperatures
kT=1.72$^{+1.23}_{-0.43}$~keV and 1.66$^{+0.53}_{-0.39}$~keV, and
contribute $\sim$6 and $\sim$5 per cent of the total flux. Freeing
elemental abundances again improves the fit without significantly altering
the primary temperature, but does not produce a well constrained fit. Fits
to bins 3 and 4 of the \xmms\ profile are not improved by additional
spectral components. We also tried fitting using the \textsc{xspec xmmpsf}
model, which corrects for the scattering of photons between spectral bins
by the broad \xmms\ PSF. Inclusion of the model introduced instabilities in
the abundance profile, with poor fits in the outer bins leading to a
``ringing'' effect wherein bins had alternately high and low abundances.
Freezing abundance at the values shown in Figure~\ref{fig.deproj} for the
four outer bins produced a more stable fit. The resulting temperature
profile was very similar to that produced without the \textsc{xmmpsf}
model, with differences in each bin smaller than the uncertainty, and
estimated uncertainties of the temperatures increasing by 5-10 per cent.
This suggests that the width of our spectral bins is sufficient to make the
spectra relatively insensitive to PSF scattering, which will be dominated
by photons scattered outward from the high surface brightness inner
regions.  We therefore believe that our deprojected temperature profile is
robust and reliable.

\subsubsection{AGN contribution}
Freezing the model fits to the outer bins in the deprojected profile, we
are able to test whether the central bin supports the inclusion of a
powerlaw component. There will be some powerlaw contribution from X-ray
binaries within NGC~3411; from the L$_{X,dscr}$:\LK\ relation of
\citet{KimFabbiano04}, we expect
L$_{X,dscr}$=5.1$\pm2.0\times10^{40}$\ergps.  In both \xmms\ and \chandra\ 
data the addition of powerlaw component produces only a small improvement
in the fit. A second plasma component produces a more significant
improvement (e.g., reduced $\chi^2$=1.26 for 111 d.o.f., compared to
1.52/112 for the powerlaw in the \chandra\ data). The powerlaw models are
poorly constrained, and in the \chandra\ fit the emission measure is
consistent with zero, within the 90\% uncertainties. The total powerlaw flux
estimated from the \xmms\ (\chandra) fit is
3.17$^{+10.12}_{-1.81}\times10^{-13}$
(6.45$^{+12.70}_{-6.45}\times10^{-13}$) \ergpspcmsq, which after
subtraction of the expected flux from X-ray binaries, gives an AGN
luminosity of \Lx=1.11$^{+5.18}_{-0.93}\times10^{41}$
(2.79$^{+6.50}_{-2.79}\times10^{41}$) \ergps.

As a further test, we extracted the projected \chandra\ spectrum from a
5\arcs\ radius circle positioned at the galaxy center. Fitting an
APEC+powerlaw model, we find a powerlaw flux of
3.16$^{+2.87}_{-1.76}\times10^{-13}$\ergpspcmsq. However, an APEC model
with variable O, Si, S and Fe provides a superior fit to the data (red.
$\chi^2$=1.308 for 46 d.o.f. compared to 1.572/48 d.o.f.).

As a final check we created 5.0-10.0~keV images for the three \xmms\ 
cameras and \chandra, and examined the area surrounding the radio source in
the optical centre of NGC~3411. No obvious source is visible in any of the
images. The \chandra\ PSF is small enough that even at 8~keV we would
expect $\sim$75\% of photons from a point source to be grouped within
1\arcs\ of the source position. The contribution from X-ray binaries or
from the cosmic background is negligible over such a small area.  No counts
were found within 2.5\arcs\ of the peak of the radio emission. A 3$\sigma$
upper limit on the emission from the AGN is equivalent to 10 counts , or
6.2$\times10^{40}$ \ergps\ in the 0.4-7.0~keV band. This places a firm
limit on the maximum X-ray luminosity of the AGN. The \xmms\ PSF is
broader, so we use regions similar to those chosen for spectral extraction.
Comparing the number of counts in a region of 30\arcs\ radius to those in a
background region at 90-120\arcs, we find no statistically significant
differences in the EPIC MOS images. The EPIC PN, which has the deepest
exposure, has a marginally significant (3.0$\sigma$) excess, with
21.2$\pm$8.7 more counts than would be expected. If we assume a powerlaw
model with $\Gamma$=1.65, this corresponds to a luminosity of
3.2$\times$10$^{40}$\ergps in the 0.4-7.0 keV band, with a 3$\sigma$ upper
limit of 7.1$\times$10$^{40}$\ergps. This is comparable to the flux
expected from X-ray binaries for the galaxy as a whole. We therefore
consider the AGN to have a maximum X-ray luminosity of
$\sim$6$\times$10$^{40}$\ergps, and that its true luminosity is likely
considerably lower.

\subsubsection{Elemental abundances}
\begin{figure}
\centerline{\includegraphics[width=\columnwidth,bbllx=30,bblly=210,bburx=560,bbury=750,clip=]{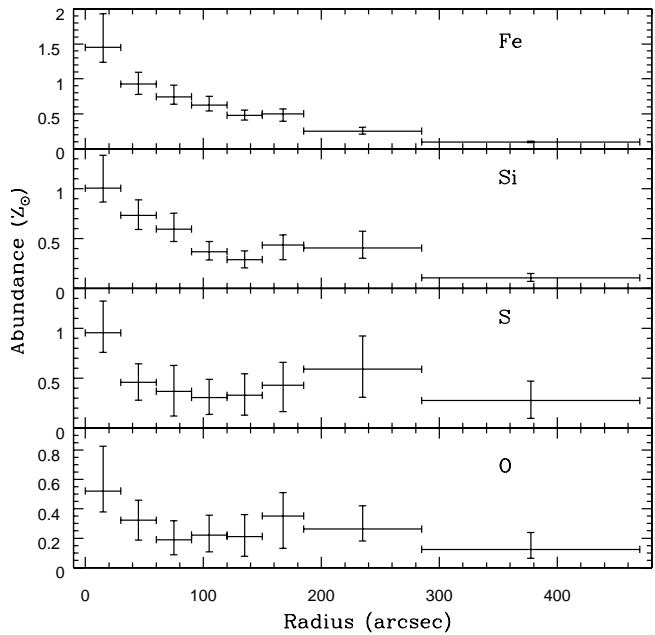}}
\caption{\label{fig.metals}
Projected abundances of O, Si, S and Fe, from fits to \xmm\ data. 90\%
uncertainties are indicated.}
\end{figure}

Allowing individual elemental abundances to vary freely in the deprojected
fits produced poorly constrained results with large uncertainties, even
when larger bin sizes and higher quality spectra were used. We therefore
chose to fit the projected elemental abundance profiles. For the central
six bins, two-temperature plasma models were found to be required to
achieve an acceptable fit. For the outer two bins only a single temperature
model was required. Figure~\ref{fig.metals} shows the resulting profiles
for 0, Si, S and Fe. The abundance of all four elements peaks in the core,
Fe having the highest abundance. Fe and Si appear to decline slowly with
radius, tracing enrichment of the gas.  S and O show a central peak with
fairly flat (though poorly constrained) abundance profiles at larger
radius, with O having the lowest central abundance of the four elements
measured. The general decline with radius is similar to that seen in poor
clusters \citep[e.g.][]{Tamuraetal04}, as is the lower O abundance. Fewer
groups have published elemental abundance profiles. Similar abundance
gradients and relative O abundances are found in NGC~5044
\citep{Buoteetal03b} and HCG~62 \citep{Moritaetal06}.  The slightly hotter
($\sim$1.5~keV) systems NGC~1399 and NGC~507 also show strong central
abundance peaks and declines in both Fe and Si
\citep{Buote02,KimFabbiano04b}, although in NGC~507 generally higher
abundances are reported ($\sim$3~\Zsol\ in the centre, dropping to
$\sim$0.7~\Zsol\ at $\sim$80~kpc). Based on the available data, we
conclude that the metallicities we measure are not unusual for a
$\sim$1~keV system such as NGC~3411.

\subsection{Spectral maps}
\label{sec.maps}
Variations of spectral properties which are not azimuthally symmetric
require a more complex spectral analysis than that described above. To look
for such variations, we created 2-dimensional spectral maps of the inner
halo of NGC~3411 using the techniques described in \citet{OSullivanetal05}.
The mapping process can be summarized as follows. A grid of square regions
is defined, covering the area of interest, each of which will be a pixel of
the final map. Square spectral regions are defined, centered on each map
pixel, and their size is allowed to vary so that each contains a minimum
number of counts. Pixels whose spectral region would be above a fixed
maximum size are excluded from the map. For maps of \xmms\ data, the most
recent ``canned'' responses are used, the software selecting the most
appropriate response files for each region. For maps based on \chandra, a
grid of response matrices is produced for the map. The spectra are fitted
in \textsc{isis} v1.3.0, and resulting parameter values (or uncertainties)
entered in the corresponding map pixel to create a map showing the
variation of that parameter across the field.

For the maps of NGC~3411, we chose to use 6.4\arcs\ pixels (\ie each pixel
is a 6.4$\times$6.4\arcs\ square) in the \xmms\ maps and 6.9\arcs\ in the
\chandra\ maps. The initial pixel grids were square, containing 4096
pixels, though a number were removed due to lack of counts. Each spectrum
was required to have at least 800 counts, and spectra were grouped to 30
counts per bin. The spectral extraction regions varied in size from
8-60\arcs\ for \xmms\ and 8.55-64\arcs\ for \chandra. It might be expected
that the \xmms\ data, which has a factor $\sim$3.4 more counts, would
produce significantly smaller box sizes than the \chandra\ data. The
broader \xmms\ PSF prevents this, by blurring the surface brightness peak;
the \chandra\ data have fewer total counts, but a higher central surface
brightness. However, while the \xmms\ data have a larger number of spectral
extraction regions of the smallest size.  The fact that the spectral
extraction regions are larger than the map pixels means that spectral fits
are not independent; all spectral regions overlap to some degree.  However,
experience with spectral maps of other systems
\citep{OSullivanetal05,OSullivanPonman04b} has shown that this technique
does a surprisingly good job of identifying temperature and abundance
structures which may be on smaller scales than the spectral extraction
regions.

\begin{figure*}
\centerline{\includegraphics[width=\textwidth]{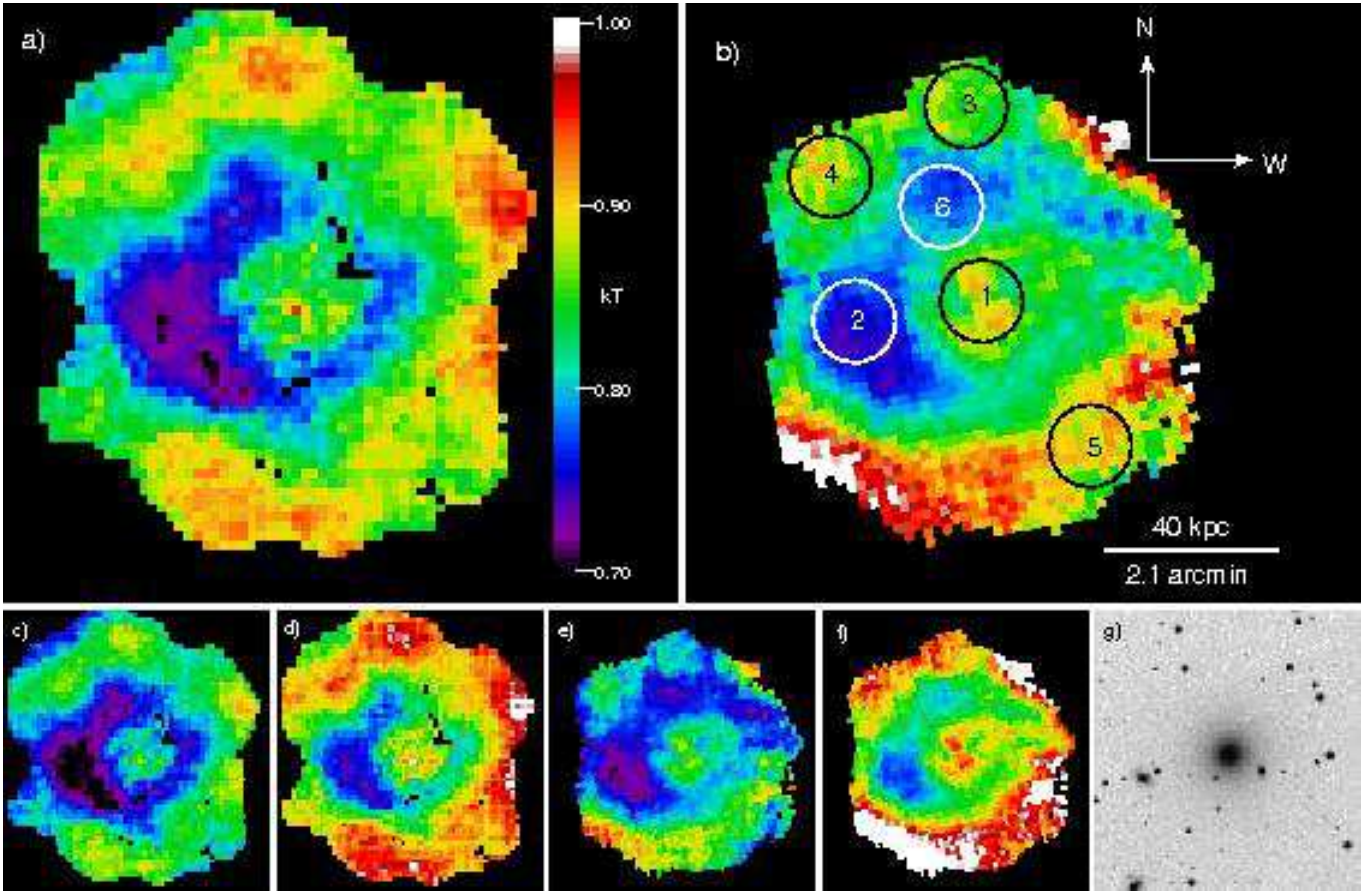}}
\caption{\label{fig.tmaps}
  Temperature maps of the inner halo of NGC~3411. Panels a) and b) show the
  best fitted temperature from the \xmms\ and \chandra\ data respectively.
  Panels c), d), e) and f) show the 90\% lower and upper bounds on
  temperature for the two datasets. Panel g) shows a DSS optical image with
  the same scale and orientation as the maps. All maps have color scales
  matching the bar in panel a). One \xmms\ map pixel is 6.4\arcs\ across,
  and one \chandra\ map pixel is $\sim$6.9\arcs. Numbered regions in panel
  b) show regions used to test the accuracy of the maps.}
\end{figure*}

Figure~\ref{fig.tmaps} shows temperature maps based on \xmm\ and \chandra\ 
data, in which a number of important features are visible. Most obvious is
that the maps can be divided in to three regions; a central roughly
circular warm core, surrounded by a more elliptical area of cooler gas,
which is in turn surrounded by warmer gas at the edge of the maps. The
temperature differences between these regions are relatively small, perhaps
0.15~keV, but are certainly $>$90\% significant. Comparison of these
regions to the radial temperature profiles discussed in
Section~\ref{sec.spectra} shows that the map temperatures correspond
closely to the higher quality spectra extracted from the radial bins.

Abundances were allowed to vary freely in the spectral fits to each map
pixel, but were not well constrained, and no clear structures were found in
the abundance distribution. Individual fits produced typical values of
0.3-1.0\Zsol. The poor constraints are unsurprising given the small numbers
of counts used in the spectra. In order to check the accuracy of the map,
we chose several 30\arcs\ radius circular regions of relatively constant
temperature, shown in Figure~\ref{fig.tmaps}. Spectra from these regions
were extracted and fitted as described in Section~\ref{sec.spectra}. A
comparison of the fitted temperatures to those in the map is shown in
Table~\ref{tab.maptest}. The two sets of temperatures are very similar, and
we therefore conclude that the maps are an accurate and reliable guide to
the temperature structure of the inner halo.

\begin{deluxetable}{lcccc}
\tablewidth{0pt}
\tablecaption{\label{tab.maptest} Accuracy of spectral map temperatures}
\tablecolumns{5}
\tablehead{\colhead{Region} & \multicolumn{2}{c}{kT (keV)} & \multicolumn{2}{c}{Map kT (keV)} \\ \colhead{} & \colhead{XMM} & \colhead{Chandra} & \colhead{XMM} & \colhead{Chandra}}
\startdata
1 & 0.84$\pm$0.004 & 0.84$\pm$0.01 & 0.81-0.95 & 0.81-0.91 \\[+1mm]
2 & 0.74$\pm$0.02 & 0.73$\pm$0.03 & 0.70-0.75 & 0.73-0.79 \\[+1mm]
3 & 0.94$\pm$0.05 & 0.88$^{+0.06}_{-0.07}$ & 0.84-0.94 & 0.82-0.89 \\[+1mm]
4 & 0.88$^{+0.03}_{-0.04}$ & 0.87$^{+0.07}_{-0.06}$ & 0.84-0.92 & 0.82-0.92 \\[+1mm]
5 & 0.88$\pm$0.05 & 0.89$\pm$0.06 & 0.84-0.91 & 0.84-0.92 \\[+1mm]
6 & 0.75$\pm$0.01 & 0.75$\pm$0.03 & 0.73-0.80 & 0.76-0.82 
\enddata
\end{deluxetable}

\begin{figure}
\centerline{\includegraphics[width=7cm]{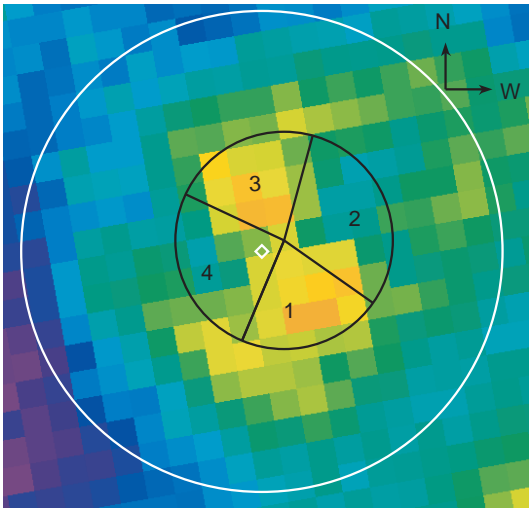}}
\caption{\label{fig.mapcentre}
  Expanded view of the central region of the \chandra\ temperature map. The
  color scale is identical to that of Figure~\ref{fig.tmaps}, and each map
  pixel is 6.9\arcs\ or $\sim$2.2~kpc across. The white circle marks the
  \Dtf\ ellipse of NGC~3411, and is $\sim$38.8~kpc across. The white
  diamond marks the optical centre of the galaxy and the peak of the radio
  emission. The four wedge regions used to examine the temperature
  variation in the central halo are marked and numbered in black.}
\end{figure}

One notable difference between the \chandra\ and \xmms\ maps is the degree
of structure within the central hot core. The \xmms\ map suggests that the
temperature in the central $\sim$1\arcm\ is fairly constant. The \chandra\ 
map appears to show hotter regions to north and south, with cooler gas to
the east and west, and the uncertainties on the map temperatures are small,
suggesting that these features may be significant.
Figure~\ref{fig.mapcentre} shows an expanded view of this region of the
\chandra\ map, with the \Dtf\ contour and optical centre of NGC~3411 marked
for comparison. To test the significance of the temperature differences, we
extracted spectra from four sector regions, marked in
Figure~\ref{fig.mapcentre}, chosen to approximately cover the two hotter (1
and 3) and cooler (2 and 4) regions. Spectral fitting confirmed small but
significant differences in fitted temperatures between the regions. Using
single spectra for each pair of regions, to maximize the signal-to-noise
ratio of the data, we find a 4.5$\sigma$ significant difference of
0.07~keV. Similar fitting of the \xmms\ data does not show any significant
difference; however, the features in question are of comparable size to the
\xmms\ PSF ($\sim$20\arcs\ across) and it seems likely that the higher
resolution of \chandra\ may be allowing us to detect features which are
blurred out by \xmms.

\subsection{Mass analysis}
\label{sec.mass}
Given the measured temperature and surface brightness profiles, it is
possible to estimate a number of other properties of the gas halo,
including the density, gas mass, entropy and cooling time profiles. Gas
density can be estimated from the measured profiles and normalized to
reproduce the X-ray luminosity of the group. The total mass profile can
then be estimated from the density and temperature profiles using the
well-known equation for hydrostatic equilibrium

\begin{equation}
M_{tot}(<r) = -\frac{kTr}{\mu m_pG}\left(\frac{d{\rm ln}\rho_{gas}}{d{\rm
      ln}r}+\frac{d{\rm ln}T}{d{\rm ln}r}\right).
\end{equation}

A simplified definition of entropy, ignoring logarithms and constants, is
used

\begin{equation}
S = \frac{kT}{n_e^{2/3}}.
\end{equation}

Uncertainties on the derived parameters are estimated using a Monte-Carlo
technique. The known uncertainties on the temperature and surface brightness
models and the total luminosity are used to randomly vary the input
parameters, and 10000 realizations of the derived parameter profiles
generated. These are then used to estimate the 1$\sigma$ uncertainty on each
parameter at any given radius.

The main issue which arises in these calculations is the question of how
well the temperature profile can be modelled. A range of models is
available, but the unusual temperature profile of the group makes an
accurate fit difficult. Specifically, no model produces an acceptable
description of both the cool shell and central peak. We therefore fit two
separate models, both of which reproduce the temperature profile from
120\arcs\ outward. In one we ignore the cool shell, fitting a model similar
to the universal cluster temperature profile described by
\citet{Allenetal01b}, with a central high temperature, outer low
temperature and a gradient in between. In the second (which we will refer
to as the cool core model), an extra term is added to the model, allowing a
central decline in temperature, which matches the cool shell but ignores
the central two temperature bins. Each model is used to generate a set of
mass, entropy and cooling time profiles, and these can be compared to
examine the differing effect of the different models.

For the mass-to-light profile, a deprojected optical luminosity profile of
NGC~3411 is calculated, using an approximation to a 3-dimensional
de Vaucouleurs density profile \citep{MellierMathez87}. An effective radius
of $r_e=30.7$\arcs\ is assumed \citep{Faberetal89} and the profile is
normalized to match the total B band luminosity of the galaxy.

\begin{figure*}
\centerline{\includegraphics[width=18cm,bbllx=30,bblly=220,bburx=590,bbury=750,clip=]{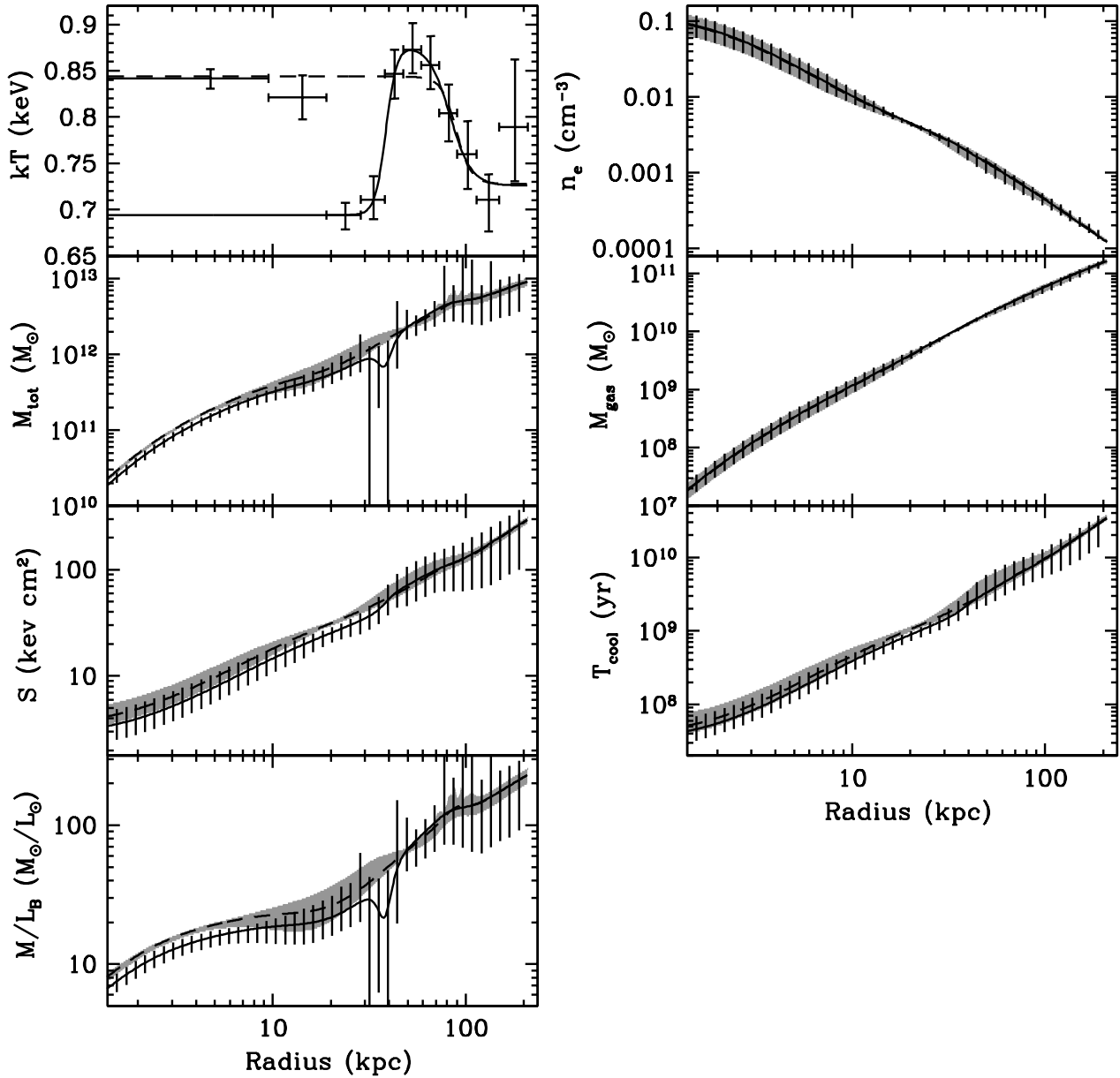}}
\caption{\label{fig.mass}
  Best fitting surface brightness and deprojected temperature models, and
  profiles of mass, entropy, cooling time and mass-to-light ratio (M/L)
  derived from them. The first panel shows the \xmms\ deprojected
  temperature profile as crosses with 1$\sigma$ uncertainties on kT. The
  dashed line and grey 1$\sigma$ uncertainty region indicate profiles
  derived from the Allen et al. style temperature model described in the
  text, in which the third and fourth temperature bins were ignored. The
  solid line and hatched 1$\sigma$ uncertainty region show profiles based
  on the cool core model discussed in the text, fitted by ignoring the two
  central temperature bins.  Note that the M/L profile includes only light
  from NGC~3411 itself.}
\end{figure*}

Figure~\ref{fig.mass} shows the resulting profiles. As would be
expected, the density and gas mass profiles are almost unaffected by the
choice of temperature model. Entropy and cooling time show only minor
differences, the cool core model producing slightly lower values but larger
uncertainties so that the two profiles are generally quite comparable. A
number of the parameters of the cool core temperature model have correlated
errors, and this increases the uncertainty on the derived mass profile,
particularly at large radii.  The main difference occurs in the total mass
profile. At the point where the temperature begins to fall in the cool core
model, a dip is visible in the total mass profile. This is of course
unphysical; total mass cannot decline with increasing radius, as this would
require the presence of material with negative mass. The dip is produced by
the very steep gradient in T at this point.  This may indicate that the
sharp gradient is a product of the bin size we have used, and that were we
able to use a higher resolution temperature profile the gradient would be
somewhat reduced. Alternatively it may indicate a deficiency in our surface
brightness model, as a steeper surface brightness gradient at that radius
would counter the temperature decline. However, it seems unlikely that
either of these factors could affect the mass profile enough to prevent the
unphysical dip in total mass.  A third possibility is that our assumptions
about the gas are incorrect. We find little evidence for multi-phase gas at
this radius, so the most likely cause is that the gas is not in hydrostatic
equilibrium. Calculating the mean enclosed mass directly from the
temperature and normalisation of the spectra in bins 4 and 5 we find that
M($<$r) does indeed rise with radius.  However, the bins are too broad to
rule out some deviation from hydrostatic equilibrium on small scales.
Nonetheless, apart from the region immediately around this feature, both
total mass and mass-to-light ratio (M/L) profiles are quite similar, within
the uncertainties, and both have physically realistic values.

\subsection{Radio structure}
\label{sec.radio}
\begin{figure}[t]
\centerline{\includegraphics[width=\columnwidth]{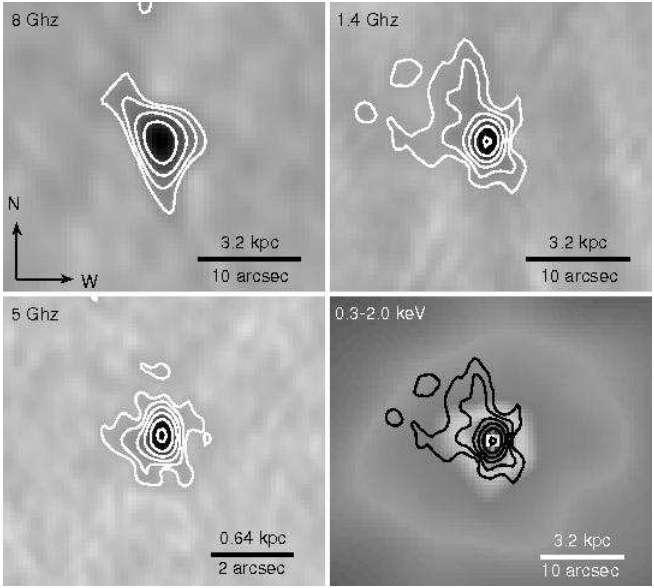}}
\caption{\label{fig.radio}
  8, 5 and 1.4 GHz VLA maps of the core of NGC~3411 with radio flux density
  contours overlaid, and an adaptively smoothed 0.3-2.0 keV \chandra\ image
  with the 1.4~GHz radio contours overlaid for comparison. Contour levels
  begin at 5$\sigma$ (0.105 mJy/beam at 8 GHz, 0.0625 mJy/beam at 5 GHz,
  0.14 mJy/beam at 1.4 GHz) and increase in factors of 2. See
  Table~\ref{tab.vlabasics} for beam sizes and R.M.S. noise in the three
  radio bands. The \chandra\ image was smoothed using \textsc{csmooth} with
  a signal-to-noise ratio range of 3 to 5.}
\end{figure}

Images from our three VLA observations are shown in Figure~\ref{fig.radio}.
The peak of the radio emission corresponds closely to both the peak of the
X-ray halo and the optical centre of NGC~3411. The NVSS data (see
Figure~\ref{fig.overlay}) show a resolved source with extensions to the
south and north-west on a 1\arcm\ scale. Our observations provide much
higher spatial resolution, and show details of the structures in the centre
of the NVSS source. Both the 5~GHz and 1.4~GHz data show a resolved nuclear
source, with a region of low surface brightness emission extending
$\sim$5\arcs\ to the north-east in the 1.4~GHz image. The beam size of the
8~GHz data is a little too large to resolve the core, but again there is
evidence of a central peak and some extension to the north-east.

\begin{deluxetable*}{lcccccc}
\tabletypesize{\footnotesize}
\tablecaption{\label{tab.radio}Parameters of model fits to the central radio component}
\tablehead{\colhead{Frequency} & \colhead{R.A.} & \colhead{Dec.} &
  \colhead{FWHM} & \colhead{P.A.} & \colhead{Peak flux density} &
  \colhead{Integrated Flux Density} \\ \colhead{(GHz)} & \colhead{} & \colhead{} & \colhead{($^{\prime\prime}$)} & \colhead{($^\circ$)} & \colhead{(mJy/beam)} & \colhead{(mJy)}}
\startdata
1.4 & 10 50 26.09 & -12 50 42.2 & 1.51$\pm$0.05$\times$1.43$\pm$0.06 & 123$^{+25}_{-31}$ & 5.07$\pm$0.15 & 8.27$\pm$0.25\\[+1mm]
5 & 10 50 26.08 & -12 50 42.3 & 0.12$^{+0.04}_{-0.05}$ & 31$^{+13}_{-12}$ & 2.18$\pm$0.07 & 2.19$\pm$0.07 \\[+1mm]
- & 10 50 26.10 & -12 50 42.4 & 1.70$^{+0.22}_{-0.25}\times1.40^{+0.23}_{-0.22}$ & 91$^{+40}_{-39}$ & 0.16$\pm$0.03 & 1.52$\pm$0.29\\[+1mm] 
8 & - & - & 2.1$\pm0.2\times0.5\pm0.5$ & 61$\pm$5 & 1.5 & 1.8 
\enddata
\end{deluxetable*}

Using the \textsc{aips imfit} task, we fitted Gaussian models to the
nuclear source visible in the 1.4 and 5~GHz images, and the best-fit
parameters of these models can be found in Table~\ref{tab.radio}. For the
5~GHz image, a single Gaussian was found to produce a poor fit, and a
second component was added. The more extended component at 5~GHz appears to
correspond fairly well with the 1.4~GHz model, both having similar extent
and rather poorly defined position angles. A Gaussian model was also fit to
the 8~GHz data, but was not as well constrained. Based on the 1.4 and 5~GHz
fluxes, the nuclear source has a spectral index $\alpha$=0.63$\pm$0.07, and
a total luminosity, integrated between 10~MHz and 10~GHz, of
L$_R$=2.7$\times10^{38}$ \ergps.

The diffuse emission seen in the 1.4~GHz image has an integrated
flux of $\sim$5.2 mJy, giving a total flux of $\sim$13.5 mJy. This is
somewhat lower than the NVSS estimate of total flux, S$_{1.4\rm{GHz}}$=33 mJy,
which may indicate that there is larger scale diffuse, low surface
brightness emission which is not detected in our observations. Several
other radio sources were found in the field of view of our observations, but
comparison with the digitized sky survey images of the field and with
NED\footnote{the NASA/IPAC Extragalactic Database,
  http://nedwww.ipac.caltech.edu} suggests that these are background
objects not associated with the NGC~3411 group. 

\section{Discussion} 
\label{sec.discuss}
The most notable feature of the NGC~3411 system is the central temperature
peak and surrounding cool shell. Similar temperature structures are seen in
a handful of systems. \chandra\ temperature profiles of NGC~1407 and
NGC~4649 show central rises of $\sim$0.2~keV within the central kiloparsec
of each group dominant galaxy \citep{Humphreyetal06b}. Both systems have
temperature profile which rise with radius, but with relatively shallow
gradients compared to the steep rise in temperature seen at the outer edge
of the cool shell in NGC~3411. Smooth mass profiles can be derived for the
two groups, so any deviations from hydrostatic equilibrium must be mild.
The cause of these central temperature rises are unclear, and both AGN
heating or merger activity are possibilities. In either case, the systems
seem to have been only weakly affected or to have had sufficient time to
become largely relaxed.  Abell 644 may be an example of a more disturbed
system \citep{Buoteetal05}.  In this cluster, the X-ray surface brightness
peak, dominant cD galaxy and centroid of the X-ray halo are all offset from
one another. If a radial temperature profile is measured centered on the
X-ray peak, the central $\sim$40~kpc region is found to be $\sim$2~keV
higher than in the surrounding bins. Measuring the profile from the
centroid of the X-ray halo removes this peak and the surrounding low
temperature region, producing a near isothermal profile. The temperature
structure is thought to be the product of a merger which has heated the
cluster core and produced a ``sloshing'' motion, leading to the observed
offsets. Using the centrally peaked temperature profile, Buote \etal\
derive a mass profile which is very uncertain, but consistent with the NFW
model \citep{NavarroFW97}. However, they argue that the central 75~kpc of
the cluster is likely to be out of hydrostatic equilibrium, given the
observed disturbance. NGC~3411 shows no evidence of ``sloshing'' or of a
significant merger capable of producing such a disturbance.

In NGC~3411, the shell of cool gas surrounds the hotter core at a radius of
$\sim$20-40~kpc. The temperature maps show that the shell is not completely
regular, and the two maps do not entirely agree on the projected
distribution of the gas. However, both show cool gas completely surrounding
the hotter core region, and both show the largest concentration and the
coolest gas to the east of the core. If we approximate the gas distribution
as a uniformly filled spherical shell, we find that it contains
9$\pm$1$\times10^9$ \Msol\ of gas. If we instead assume a shell only 10~kpc
thick, with an additional ``blob'' of gas of size 40$\times30\times20$~kpc
(corresponding roughly to the region east of the core), the mass of cool
gas is reduced to $\sim$3.9$\times10^9$ \Msol.  Neither model is physically
rigorous; the most likely true configuration is a spherical hot core
located towards one end of a slightly elliptical cool shell. For
simplicity, we will assume the former approximation in our discussion, but
note that the mass and volume may be a factor $\sim$2 lower.

\subsection{Shell formation by AGN reheating of a cool core}

Perhaps the most obvious scenario for the formation of the peak and shell
is that NGC~3411 is a system in which the center of a cooling region has
been reheated. In this model, the group halo would have been undisturbed
until relatively recently, and radiative cooling would have produced a cool
inner region with a smoothly declining central temperature profile. Cooling
of gas in the core of NGC~3411 itself could then provide fuel for an AGN
outburst (or potentially for star formation, though the current absorption
dominated optical spectrum argues against this) which would provide enough
energy to partially reheat the gas, producing the hot core which we
observe. We can estimate the minimum energy required by assuming that the
core was heated from a temperature of 0.7~keV (equal to the cool shell) to
its current state.  This would require $\sim2\times10^{57}$\erg. Using the
relation between AGN luminosity and mechanical power output
\citep{Birzanetal04}, we can estimate the current rate of energy injection
from the AGN in NGC~3411 to be $\sim$8.2$\times10^{41}$\ergps.  This is
somewhat higher than the upper limits on the current X-ray luminosity of
the AGN, but is considerably lower than the rate at which energy is lost
from the core through radiative cooling, 1.13$\times10^{42}$ \ergps. This
strongly suggests that if AGN heating is responsible for the raised
temperature of the core, the AGN power output must have been significantly
larger in the past. Energy injection at 10$^{43}$\ergps\ could produce this
change in $\sim$7.2 Myr, while injection at 4 times the current mechanical
power estimate would take $\sim$30~Myr. However, these energy requirements
should be considered lower limits; if the group did possess a cool core
prior to reheating, the centre of that region would likely have been cooler
than 0.7~keV. The maximum energy injection rate is limited by the fact that
we see no sign of convective instability, which would be expected if the
heating were very rapid. The smooth entropy gradient observed shows that
the gas heating rate was not sufficient to drive convective gas flows.

We can also calculate the energy required to create the two ``hot spots''
within the core region. As we are not sure of the mechanism by which
heating has occurred, we use the local sound speed, $\sim$365 \kmps, to
estimate a rough timescale of 15.5~Myr over which the hot regions formed.
Approximating each hot region as a sphere $\sim$6.5~kpc in diameter and
assuming that they were heated by 0.07 keV, we find that the energy
required is $\sim$1.7$\times10^{55}$ erg, equivalent to 8$\times10^{40}$
\ergps\ over the estimated timescale. This is a relatively low power
compared to that estimated for the hot core as a whole. 

The fact that we see no signs of jets or shocks in the radio or X-ray data
suggests that the AGN is currently in a fairly quiescent state, or that it
is injecting energy through some form of outflow which is not radio or
X-ray bright. Comparison with low-luminosity AGN \citep{TerashimaWilson03}
and weak-line radio galaxies \citep{Rinnetal05}, and with the black hole
fundamental plane \citep[using a black hole mass of
M$_{SMBH}\sim4\times10^8$~\Msol, based on galaxy
luminosity][]{HaringRix04,Merlonietal06}, suggests that NGC~3411 may be a
fairly typical low-luminosity system, though it lies at the low end of the
range of both X-ray and radio power. However, if the X-ray luminosity is
significantly lower than out upper limit, the AGN would be classed as
uncommonly X-ray faint.  

Some Seyfert galaxies and broad-absorption-line (BAL) QSOs are known to
have diffuse bipolar radio structures centered on the AGN, similar to the
radio structure we see in the core of NGC~3411
\citep{Stockeetal92,UlvestadWilson84}. Such structures may represent
outflowing winds from the accretion disk, driven by radiation pressure or
magnetocentrifugal forces \citep[][and references
therein]{Elvis00,DeKool97}. However, numerical modelling of such winds
\citep{Proga03} and measurements from nearby systems
\citep[e.g.][]{Kraemeretal05} suggest that the energy output from such a
wind would be rather low. NGC~3411 has a considerably more massive SMBH
than a typical Seyfert galaxy and is perhaps more comparable with BAL QSOs,
but the output would still likely be only a few 10$^{38-39}$~\ergps, to low
to be responsible for heating or maintaining the temperature in the group
core.  In any case, the lack of a definite X-ray source associated with the
radio peak, the substructures to the north and south of the radio core, the
lack of a decrease in central surface brightness (as would be expected if a
jet in the line of sight had excavated cavities in the X-ray halo) and the
absorption dominated optical spectrum all strongly suggest that the AGN is
heavily absorbed, and probably has an axis close to the plane of the sky.

\subsection{Shell formation from stripped gas}

An alternative is that the cool shell is the product of a merger. It is
well established that galaxies falling through the IGM of a galaxy cluster
can be stripped of their gas by ram pressure, with several examples
observed in nearby systems
\citep[e.g.,][]{Machaceketal06,Machaceketal05a,Davisetal97}. An early-type
galaxy, falling into the NGC~3411 group and passing through the core, could
potentially lose a significant fraction of its gaseous halo. The stripped
gas would sink or rise to a level where its entropy matches that of the
surrounding IGM, and spread along the equal entropy surface, forming a
shell. As the stripped gas would have a fairly constant entropy, we might
expect the entropy profile to flatten at the radius of the cool shell, and
to some extent this is visible in our entropy models.  The 0.7~keV
temperature observed in the shell is quite typical of both group and
elliptical galaxy halos. If the shell is the product of a merger, no energy
input from the AGN would be required, except perhaps to balance energy lost
through radiative cooling.

There are however a number of potential problems with this model.  The
smoothly declining abundance profile shows no feature at the position of
the cool shell, arguing that the gas has the same origin and enrichment
history as the warmer regions around it. Ram-pressure stripping is most
effective at high velocities and in high density gas . Galaxy clusters
often have cores of radius $\sim$100~kpc, within which the density
n$_e\sim$0.01\pcmcu\ \citep[e.g.,][]{SandersonPonman03}. The NGC~3411 halo
reaches comparable densities only in the central $\sim$20~kpc, so an almost
radial orbit would be required. Turbulent and viscous stripping mechanisms
can be effective at lower velocities, but these are less efficient at
removing gas \citep{Nulsen82}, suggesting that the cool shell would
represent only a part of the halo of the stripped object. The amount of gas
in the shell is quite sizable. \citet{Athey03} measured the relation
between halo gas mass and optical luminosity in elliptical galaxies, and we
can use this to predict the size of a galaxy which could produce the shell.
If we assume that all the gas in the shell is from the stripped galaxy, and
that the galaxy halo was entirely removed, the relation suggests that we
should expect an object with log \LB=11.2 \LBsol.  The scatter in the
relation is quite large, but this still suggests a galaxy of comparable
or greater optical luminosity than NGC~3411 itself. The next most luminous
galaxy in the group is a factor $\sim$1.5 fainter. 

To identify galaxies which might have interacted with NGC~3411 at some
point in the past, we selected those objects within 2000\kmps\ and 1~Mpc on
the sky. Comparing K-band luminosities (which avoid bias from young stellar
populations), we find that there are four nearby galaxies of comparable
size; NGC~3404 (\LK=11.24\LKsol), IC~647 (10.32), NGC~3421 (10.83),
NGC~3422 (10.96).  All four galaxies have recession velocities within
250\kmps of that of NGC~3411. Of these, the location of the S0 IC~647
inside the cool shell just to the east of NGC~3411 might initially suggest
a connection.  However, for its halo to be the source of the gas in the
cool shell, it would have to have been stripped during a previous passage
through the core so the observed superposition would be a very unlikely
conspiracy.  It may be that the galaxy halo (if one exists) affects the
temperature map, decreasing the temperature at that position.  However,
while there is excess surface brightness in the region of IC~647 we cannot
say how much is associated with the galaxy. NGC~3421 is a disturbed face-on
spiral galaxy $\sim$0.5~Mpc to the north of NGC~3411. While late-type spirals
predominantly contain cool gas, it has been suggested that vigorous star
formation may heat this gas and cause it to expand and diffuse away from
the galactic disk, making ram-pressure stripping more efficient
\citep{Rasmussenetal06}. However, NGC~3421 has several regions of ongoing
star formation, casting doubt on the possibility that it has been
significantly stripped. The two remaining galaxies, NGC~3422 and NGC~3404
are distant enough that the cool shell would have to be unrealistically old
(1-2~Gyr) for them to have been its source, assuming motion at the group
velocity dispersion of 500~\kmps.

\subsection{Conduction across the shell boundaries}

We can put a lower limit on the lifetime of the cool gas shell by
estimating the rate at which it should be heated by conduction from the
surrounding gas. We again assume a uniform shell of gas, 20~kpc thick, and
that the process does not affect the surrounding hot gas in any way. We
assume a temperature gradient of 0.14~keV over 9.5~kpc, the distance
between the mid-points of the \xmms\ spectral bins. The heat flux across
this gradient can be described as \citep{Spitzer62,Sarazin88}:

\begin{equation}
q=\kappa\frac{{\rm d}(kT_e)}{{\rm d}r},
\end{equation}

where $q$ is the heat flux, $T_e$ is the electron temperature and $\kappa$
is the thermal conductivity, defined as \citep{CowieMcKee77}:

\begin{equation}
\kappa=1.31n_e\lambda_e\left(\frac{kT_e}{m_e}\right)^{1/2},
\end{equation}

where $n_e$ is the electron density, $m_e$ is the electron mass, and
$\lambda_e$ is the electron mean free path. In a fully ionized gas mainly
composed of hydrogen, the mean free path is defined as
\citep{EttoriFabian00}:

\begin{equation}
\lambda_e=30.2T^2n^{-1}\left(\frac{{\rm ln}\,\Lambda}{37.9+{\rm ln}(T/n^{1/2})}\right)^{-1} {\rm kpc},
\end{equation}

where $T=kT_e$/10~keV, $n=n_e/10^{-3}$\pcmcu, and $\Lambda$ is the impact
parameter of the Coulomb collisions, defined as:

\begin{equation}
{\rm ln}\,\Lambda=29.7+{\rm ln}\,[n_e^{-1/2}(T_e/{\rm 10^6 K})].
\end{equation}

From these equations, we find a mean free path of $\sim$54~pc, and estimate
that with conduction at the Spitzer rate, the heat flux across the
boundaries of the shell would be 1.96$\times10^{43}$\ergps, suggesting that
the shell would be heated to the surrounding temperature in $\sim$8~Myr.
This value would be reduced if the temperature gradient were steeper, but
would increase if the hot and cool regions were magnetically separated. In
undisturbed regions of the IGM, magnetic field lines are commonly assumed
to be tangled, producing only small reductions of the conduction rate.
However, processes which introduce structure into the IGM (\eg radio jets
and lobe inflation, mergers, and gas motions) are thought to straighten the
field lines, suppressing conduction across boundaries by large factors. In
the case of NGC~3411, the amount of suppression required depends on our
model for the formation of the cool region. If it is a cool core which has
been reheated by AGN activity, only mild suppression is necessary, as the
timescale to heat the core is $\sim$30~Myr. Heating and expansion of the
gas in the central region could cause suppression along the inner shell
surface, as field lines are stretched, and reducing conduction by a factor
of $\sim$20 could be sufficient. If the shell is stripped gas, it must be a
much more long-lived structure, with an age of at least a few 10$^8$ yr. In
this case, conduction would have to be suppressed by a factor of a hundred
or more.

One of the stronger observational constraints on conduction in galaxy
clusters comes from the temperature variations within the disturbed cluster
A754. \citet{Markevitchetal03} suggest that conduction cannot be more than
a few percent of the Spitzer rate in this system.  Studies of conduction as
a mechanism for balancing cooling in galaxy clusters suggest that rates of
a few hundredths to a few tenths of the Spitzer value are likely typical
\citep{Popeetal06,VoigtFabian04,ZakamskaNarayan03}. While these studies
tend to focus on much hotter, higher mass systems than NGC~3411, an
examination of the $\sim$2~keV Virgo cluster finds similar conduction rates
\citep{Popeetal05}. If these estimates hold true for NGC~3411, they suggest
that the conduction rate could easily be low enough to agree with our AGN
heating hypothesis. The suppression factor required by the merger model is
at the more extreme end of the range, and suggests that the other galaxies
in the group may not be viable sources of stripped gas unless their
velocity in the plane of the sky is large.

\subsection{Mass, entropy and cooling time}
The gas mass and cooling time profiles for the group halo are both rather
typical of a system of this temperature. It is notable that despite its
lower temperature, the cool shell surrounding the central elliptical still
has a longer cooling time than the slightly hotter core. The cooling time
drops below 1~Gyr at $\sim$30~kpc, around the midpoint of the cool shell,
and is 400-500~Myr at 10~kpc. This emphasizes the need for a long-term heat
source to prevent runaway cooling. As discussed in section~\ref{sec.mass},
the total mass profile is unusual only in the unphysical decline at
$\sim$40~kpc, which seems to indicate a deviation from hydrostatic
equilibrium at the outer edge of the cool shell. Depending on the formation
mechanism it is possible that all the gas in the shell, or even the entire
core and shell, could be out of hydrostatic equilibrium. However, as the
mass profile only becomes unphysical at the outer boundary of the shell, it
seems likely that it is only in this region that the deviation from
equilibrium becomes severe, and that at smaller radii the mass profile is
still reasonably reliable. The mass-to-light ratio profile stays above the
expected stellar M/L of $\sim$5 throughout, suggesting the presence of dark
matter even at relatively small radii. Simulations of group formation
suggest that we should expect dark matter halos associated both with the
member galaxies and with the group potential as a whole
\citep[e.g.,][]{Barnes89}; the increase in the slope of M/L outside
$\sim$20~kpc seems likely to indicate the point at which the group dark
matter halo begins to dominate over the mass contribution associated with
NGC~3411.

The measured entropy profile compares reasonably well with the results of
\citet{Mahdavietal05}, unsurprisingly given that both studies make use of
the same \xmms\ dataset. There is some discrepancy at the largest radii,
with the Mahdavi et al. result being a factor of $\sim$3 lower than our
best fit value at $\sim$200~kpc. However, the two values appear to be
comparable given the uncertainties on our entropy profile and the small
amount of scatter in the Mahdavi et al data points. There may be a larger
discrepancy between our total mass and the optical estimate of
\citet{Ramellaetal02}, who suggest a mass of $\sim$2.7$\times10^{13}$
\Msol\ within a Virial radius of $\sim$1.2~Mpc. A direct comparison would
require a large extrapolation of our data, which only extend to
$\sim$200~kpc, but our total mass in that radius, $\sim$8.5$\times10^{12}$
\Msol, is high enough to suggest that the total mass of the system would be
considerably higher.  However, as the Ramella et al. estimate is based on
velocities from only five group member galaxies, the uncertainty on this
estimate is likely large, and the discrepancy may not be serious.

\section{Conclusions}
\label{sec.conc}
In order to examine the properties of the NGC~3411 group and the
interactions between the central dominant elliptical and the group halo, we
have analysed \xmm, \chandra\ and VLA observations of the
system. Our results can be summarized as follows:

\begin{enumerate}
\item The group halo has a highly unusual temperature structure. The radial
  temperature profile rises from $\sim$0.72~keV at large radius to a peak
  of $\sim$0.84~keV at $\sim$50~kpc. This hot gas encloses a shell of cool,
  $\sim$0.7~keV gas which itself surrounds the hotter $\sim$0.84~keV core.
  While these temperature differences are small, they are statistically
  significant. Spectral mapping confirms that the cool gas completely
  surrounds the core (in projection) and shows the largest concentration of
  the coolest gas to be located west of the group core.
\item The \xmm\ X-ray surface brightness profile requires two beta models
  to produce a good fit, the second component producing a slight excess at
  $\sim$30~kpc radius, corresponding to the shell of cool gas. \chandra\ 
  surface brightness modelling shows a third component in the core which is
  not resolved by the \xmms\ data. This component is extended, and is
  therefore more likely related to diffuse or unresolved emission within
  the central galaxy than with AGN activity.
\item The deprojected abundance profile is typical of a galaxy group, with
  a central peak or 1-1.5\Zsol\ and decline with radius. The profile shows
  no features corresponding to those in the temperature profile.
\item Calculation of the total mass, entropy, cooling time and
  mass-to-light ratio shows them to be fairly typical of galaxy
  groups. There is no entropy inversion in the core, indicating that
  any heating in the system has been slow. The cooling time of the system
  is shorter than 10$^9$ yr within $\sim$20~kpc, suggesting that some source
  of heating is necessary in the long term to prevent runaway cooling. The
  Mass-to-light ratio profile supports the presence of dark matter at
  all radii with the possible exception of the innermost galaxy core. A
  change in gradient at $\sim$20~kpc may indicate the point at which the
  group dark matter halo begins to dominate the mass distribution. However,
  an unphysical decline in total mass at $\sim$35~kpc suggests that the gas
  is not in hydrostatic equilibrium at this point.
\item Analysis of the VLA observations at 1.4, 5 and 8~GHz reveals a
  resolved nuclear source with a luminosity of 2.7$\times10^{38}$\ergps.
  There is also evidence for some diffuse emission at 1.4~GHz, extending a
  few arcseconds. There is no sign of well defined radio jets or lobes, and
  this may indicate that the AGN is is a relatively quiescent state. An AGN
  driven wind may be responsible for the diffuse radio emission we observe.
  There is little evidence for significant AGN activity in the X-ray data,
  and the \chandra\ data places a firm upper limit on the AGN 0.4-7.0~keV
  luminosity of $\sim$6$\times10^{40}$ \ergps. This suggests that the AGN
  is heavily absorbed, and probably has an axis close to the plane of the
  sky.
\item One possible explanation of the cool gas observed in the temperature
  analysis is that it is stripped material which has settled along its
  equal entropy surface to form a shell. However, the mass of gas is
  relatively large, stripping mechanisms are thought to be relatively
  inefficient in the group environment, and the galaxies which might be the
  source of the material are both distant and unlikely to have possessed
  sufficient hot gas. It is also possible that the timescale over which
  conduction would heat the cool shell is short enough that the other
  galaxies in the group would be unrealistic as sources of the stripped gas.
\item An alternative is that the temperature features represent a cool core
  which has been partially reheated by AGN activity. In its current
  quiescent state, the AGN is unlikely to provide the energy required for
  such heating, therefore it seems probable that it underwent a period of
  enhanced activity in the recent past. If this explanation is accurate, we
  are observing the group at a very unusual period in its history, when the
  effects of both cooling an heating are visible. The lack of other systems
  with similar temperature structure would then be unsurprising. In our
  opinion, this is the more likely scenario.
\end{enumerate}

\acknowledgments This research has made use of the NASA/IPAC Extragalactic
Database (NED) which is operated by the Jet Propulsion Laboratory,
California Institute of Technology, under contract with the National
Aeronautics and Space Administration. We also acknowledge the use of NASA's
\textit{SkyView} facility (http://skyview.gsfc.nasa.gov) located at NASA
Goddard Space Flight Center. EOS and JMV acknowledge support for this work
provided by the National Aeronautics and Space Administration through
Chandra Awards Number AR4-5012X and G02-3186X issued by the Chandra X-ray
Observatory Center, which is operated by the Smithsonian Astrophysical
Observatory for and on behalf of the National Aeronautics Space
Administration under contract NAS8-03060. We are grateful to J.~Kempner for
his assistance with spectral mapping software, and to F.~Nicastro and
A.~Zezas for interesting discussions of the AGN properties of the
system. We also thank the anonymous referee for their useful comments.

\end{document}